\input epsf.sty
\documentstyle[prd,preprint,aps]{revtex}

\def\gtsima{$\; \buildrel > \over \sim \;$}
\def\ltsima{$\; \buildrel < \over \sim \;$}
\def\prosima{$\; \buildrel \propto \over \sim \;$}
\def\gsim{\lower.5ex\hbox{\gtsima}}
\def\lsim{\lower.5ex\hbox{\ltsima}}
\def\simgt{\lower.5ex\hbox{\gtsima}}
\def\simlt{\lower.5ex\hbox{\ltsima}}
\def\simpr{\lower.5ex\hbox{\prosima}}

\tighten
\begin{document}
\headheight=70pt

\title{Radiation Recoil from Highly Distorted Black Holes}

\author{Steven Brandt${}^{(1,2)}$, Peter Anninos${}^{(3,4)}$}

\address{
${}^{(1)}$ Max-Planck-Institut f\"ur Gravitationsphysik,
Schlaatzweg 1, 14473 Potsdam GERMANY \\
${}^{(2)}$ Current address: Pennsylvania State University,
525 Davey Lab, University Park, 16802 \\
${}^{(3)}$ National Center for Supercomputing Applications,
Beckman Institute, 405 N. Mathews Avenue, Urbana, Illinois, 61801 \\
${}^{(4)}$ Current address: University of California,
Lawrence Livermore National Laboratory,
Livermore, CA 94550 \\
}

\date{\today}
\maketitle

\begin{abstract}
We present results from numerical evolutions of
single black holes distorted by axisymmetric, but
equatorially asymmetric, gravitational (Brill) waves.
Net radiated energies, apparent horizon embeddings,
and recoil velocities are shown
for a range of Brill wave parameters, including both
even and odd parity distortions of Schwarzschild black holes.
We find that a wave packet initially concentrated
on the black hole throat, a likely model 
also for highly asymmetric stellar
collapse and late stage binary mergers, can generate a maximum
recoil velocity of about 150 (23) km/sec for even (odd)
parity perturbations, significantly less than that
required to eject black holes from galactic cores.
\end{abstract}



\section{Introduction}
\label{sec:introduction}

Many popular models of active galactic nuclei, quasars and
even archetypical galaxies rely on the relativistic influence of
black holes on the surrounding environment to provide power sources
for observed spectral emissions
and inferred motions of gaseous or stellar material.
However, due to the strong gravitational effects of black holes,
their role in the evolution of
galactic cores and quasars is uncertain since numerical
relativity computations are needed
to perform detailed investigations of the near--field regime.
In particular, gravitational waves generated from
sufficiently asymmetric systems
(such as collapsing stellar cores and coalescing black holes)
can carry a nonzero linear momentum component and impart a recoil
velocity to the emitting objects,
dominantly from
the interplay between mass--quadrupole and mass--octupole
or mass quadrupole and current--quadrupole
contributions \cite{Bonnor,Perez}.
These velocities
would be astrophysically significant if they were large
enough to eject the emitting objects from the center of
the host galaxy and send them
hurtling through intergalactic space.
Because the efficiency of momentum
radiation emission is not known precisely, the 
dynamics and stability of
systems containing black hole engines remain important
but unresolved issues. 
If radiation reaction effects are significant, they may
have considerable observable consequences for astrophysics and
cosmology, including the redistribution
and depletion of black holes from host galaxies,
the disruption of active galactic core energetics,
the introduction of black holes and stellar material into the 
intergalactic medium, and the general
formation and structure attributes of galaxies.

Although approximation studies of
radiation recoil have been performed for more than
two decades now, results from these calculations based on
quasi--Newtonian and relativistic perturbation formalisms
\cite{Bekenstein,Moncrief,Fitchett1,Fitchett2,Nakamura,Andrade}
present an uncertain picture due to their incomplete treatment.
More recently, Anninos and Brandt \cite{Anninos1}
have numerically computed the
recoil effect from fully general relativistic head--on 
collisions of two 
unequal mass black holes with time--symmetric initial data,
and have shown that recoil velocities are of order
10 -- 20 km/sec for black holes with 
moderately large initial separations
($\simgt 10 M$, where $M$ is the mass of the larger black hole).
Their results are in rough agreement with,
and generally confirm, various estimates
from perturbation calculations.

Here we continue to explore the radiation reaction
process by computing the energies and recoil velocities
from single black holes distorted by 
axisymmetric gravitational (Brill \cite{Brill}) waves. 
We extend previous Brill wave + black hole
investigations \cite{Bernstein,Brandt_ivp} by
relaxing the equatorial mirror symmetry imposed in earlier work,
thus allowing for mixtures of consecutive (even/odd) multipole
contributions to the emitted radiation.
In addition to investigating radiation
reactions in this new class of physical systems,
our implementation of Brill waves allows for very highly
distorted black holes which can be thought of as models for the
late stage behavior of binary coalescing black holes. In fact,
we have been able to simulate single black hole 
distortions \cite{Brandt_ivp,Anninos_ah},
as characterized by the ratio of polar to equatorial circumferences
of apparent horizons, that are significantly greater
than what we have observed in the merged state
of two colliding black holes \cite{Anninos_ah,Anninos_2bh1,Anninos_2bh2}.
Although the horizon distortion is not the only factor influencing
recoil efficiency, one might nevertheless
expect to obtain some idea, or perhaps 
even an absolute upper limit, of the 
recoil magnitude during the late stages of
binary interactions by investigating strongly distorted
single black hole systems.

We generalize the prescription 
developed in references \cite{Bernstein,Brandt_ivp}
(and summarized for convenience in \S\ref{sec:initial})
to specify equatorially asymmetric 
initial data and to parameterize Brill
wave perturbations of Schwarzschild black holes by the amplitude,
shape, location, and spectral mixture of the even and odd $\ell$--modes.
Results from numerical evolutions are presented in
\S\ref{sec:results}, where we show embeddings of the
black hole apparent horizons, energies emitted 
in the most dominant quasinormal modes of the final black hole,
and recoil velocities arising from the mixing of consecutive radiative modes.
The computations are carried out for both even and odd parity
distortions of black holes, and over a wide range of wave strengths,
initial placements, and mode distributions.
We conclude in \S\ref{sec:summary}.

\section{Initial Data}
\label{sec:initial}

For even parity distortions, we
utilize the conformally flat approach
of Bowen \& York \cite{Bowen} to solve the initial value problem
in axisymmetry
and write the spatial 3--metric at the initial time as
\begin{equation}
dl^2 = \Psi^4\left[ e^{2(q-q_0)}\left(d\eta^2 + d\theta^2\right)
     + \sin^2\theta~d\phi^2 \right] ,
\label{eqn:metric_even}
\end{equation}
where $\Psi(\eta,\theta)$ is the conformal factor,
$q(\eta,\theta)$ is a function subject to certain constraints
in its form but is otherwise freely specifiable,
$q_0(\eta,\theta)$ is chosen so
that the Kerr metric is recovered if $q=0$ and the appropriate
extrinsic curvature is specified, $\eta$ is a
logarithmic radial coordinate centered on the black hole throat,
and $(\theta, \phi)$ are the usual
angular coordinates. 
The more general 3--metric (applicable to both even and odd 
parity perturbations) is of the form
\begin{equation}
\gamma_{ij} = \Psi^{4} \left [
\begin{array}{ccc}
A(\eta, \theta)   & 0            & 0 \\
0   & B(\eta, \theta)            & F(\eta, \theta) \sin\theta \\
0   & F(\eta, \theta) \sin\theta & D(\eta, \theta) \sin^2\theta \\
\end{array}
\right ] ,
\label{eqn:gij}
\end{equation}
with $F=0$ ($\ne 0$) for the even (odd) parity cases.
The remaining metric components ($\gamma_{\eta\theta}$ and
$\gamma_{\eta\phi}$) are set to zero by the gauge freedom in
choosing the shift vector.

As in Ref. \cite{Brandt_ivp}, the somewhat arbitrary function $q$ 
is restricted by symmetry conditions on the throat and axis,
and fall--off rates at large radii \cite{Bernstein}.
The function $q$ is constructed
to have an inversion symmetric Gaussian part given by
\begin{equation}
q = Q_0 f(\theta) \left(e^{-s_+} + e^{-s_-}\right) + q_1,
\label{eqn:q_brill}
\end{equation}
where
\begin{equation}
s_{\pm} = \frac{(\eta \pm \eta_0)^2}{\sigma^2} ,
\end{equation}
and $q_1=0$ or $q_0$ for perturbations of the
stationary Kerr solution or the Bowen \& York spacetime \cite{Bowen}
respectively. With this form,
the Brill waves are characterized by their
amplitude $Q_0$, width $\sigma$, center coordinate location $\eta_0$,
and their angular dependence $f(\theta)$.
This allows a convenient way to parameterize the
strength, shape and placement of the waves, and to
easily tune the wave data for a broad
range of spectral mode mixtures.

In previous work, we had considered the case 
$f=\sin^n\theta$ which possesses mirror symmetry across the
equator. However, this data does not allow
for the emitted gravitational waves to carry any linear momentum,
as it excludes odd multipole components.
Here, we relax the constraint of equatorial symmetry and
consider
\begin{equation}
f(n,\xi,\theta) = (1-\xi+\xi \cos\theta) \sin^n\theta ,
\label{eqn:f_pert}
\end{equation}
for which $q$ in (\ref{eqn:q_brill})
obeys the isometry conditions ($\eta\rightarrow-\eta$,
$\theta\rightarrow-\theta$, and $\theta\rightarrow 2\pi-\theta$).
The form of (\ref{eqn:f_pert}) also has
the necessary property $f(0)=f(\pi)=0$,
and regulates the even and odd mode power distributions
through the parameters $n$ and $\xi$.
The parameter $\xi$ determines the asymmetry of the wave, and the
relative excitation of the odd and 
even numbered, even parity $\ell$ modes.  When
$\xi=0$ and $n=2$, $\ell=2$ is the dominant mode; when
$\xi=1$ and $n=2$, $\ell=3$ is the dominant mode.  For
some intermediate value of $\xi$ there will be a roughly even
distribution of energy between the $\ell=2$ and $\ell=3$ modes, and
at this value the gravity waves will produce a maximum recoil
velocity on the black hole, as demonstrated in \S\ref{sec:results}. 
The initial value problem for the even parity cases is completed
by solving the Hamiltonian constraint for the conformal factor $\Psi$
in the metric (\ref{eqn:metric_even}) with the specified free data.
We also impose time symmetry, hence the extrinsic curvature
is set to zero and the momentum constraint is trivially satisfied.
This implies that the Brill wave packet is a combination
of ingoing and outgoing radiation.

For odd parity distortions, the free data is specified in the only
nontrivial momentum constraint equation arising from the $\phi$ component 
in ``time--rotation'' symmetry and maximal slicing ($tr~K=0$).
Defining the initial extrinsic curvature as
\begin{equation}
K_{ij} = \Psi^{-2} \left [
\begin{array}{ccc}
0                       & 0                   & \hat H_E \sin^2\theta \\
0                       & 0                   & \hat H_F \sin\theta \\
\hat H_E \sin^2\theta   & \hat H_F \sin\theta & 0 \\
\end{array}
\right ] ,
\label{eqn:kij}
\end{equation}
the momentum constraint reduces to
\begin{equation}
\partial_\eta(\hat H_E \sin^3\theta) +
\partial_\theta(\hat H_F \sin^2\theta) = 0 ,
\label{eqn:mom}
\end{equation}
and is satisfied by
\begin{eqnarray}
\hat H_E &=& f_1(\theta) + f_2(\eta)\left[ 4\cos\theta~f_3(\theta)
          + \sin\theta~\partial_\theta f_3(\theta) \right] , \\
\hat H_F &=& -\partial_\eta f_2(\eta) \sin^2\theta~f_3(\theta) ,
\end{eqnarray}
where $f_1$, $f_2$ and $f_3$ are arbitrary functions with
the following symmetries:
$f_1(\theta) =  f_1(-\theta) = f_1(2 \pi-\theta)$,
$f_3(\theta) =  f_3(-\theta) = f_3(2 \pi-\theta)$,
$f_2(\eta)   =  f_2(-\eta)$,
and $f_2 \rightarrow 0$ as $\eta \rightarrow \infty$
so the spacetime asymptotically approaches the Kerr solution.
To construct a model for odd parity waves analogous
to the even parity case described above, we choose the following
free functions:
\begin{eqnarray}
f_1 &=& 0 ,\\
f_2 &=& Q_0 \left(e^{-s_+}+e^{-s_-}\right) , \\
f_3 &=& \left(1-\xi+\xi \cos\theta\right) \sin^n\theta .
\end{eqnarray}
Since $\hat H_E$ falls off sufficiently rapidly at large radii,
any spacetime constructed using this conformal
extrinsic curvature with $f_1=0$ will have zero angular momentum.
The Hamiltonian constraint is then solved for $\Psi$ given
the above extrinsic curvature and a conformally flat
3-metric with $q=q_0=0$ in (\ref{eqn:metric_even}).

\section{Results}
\label{sec:results}

In this section we present results from several dozen
numerical calculations of both even and odd parity axisymmetric 
distortions of single Schwarzschild black
holes using the initial data parameterization
described in \S\ref{sec:initial}. 
The results are presented as functions of various key
Brill wave parameters, and their effect on radiation
recoil is evaluated.
In most cases we have used a numerical grid
with 300$\times$65 (radial$\times$angular) zones
to cover radial distances out to several hundred $M_{ADM}$, where
$M_{ADM}$ is the ADM mass of the spacetime, and to include
the entire polar domain $0\le\theta\le\pi$. However,
we have also confirmed that the results are robust and relatively
unchanged at different grid resolutions. Our simulations
utilize the maximal slicing condition ($K=\dot K = 0$) and are
generally run to 50 -- 70$M_{ADM}$, which is more
than enough time to extract the radiation content.
The radiated
wave energies and recoil velocities are computed from the 
energy--momentum flux across
a spherical shell of radius
15$M_{ADM}$ from the center of the black hole throat.

\subsection{Even Parity}
\label{subsec:even_schw}

First we consider the effects of varying three independent
parameters $Q_0$, $\eta_0$ and $\xi$, corresponding to
the amplitude, peak location, and dominant mode of the Brill waves,
on the radiation reaction and dynamical evolution of even parity distortions
of black holes. The remaining
free initial data parameters described in \S\ref{sec:initial}
have been held fixed: $\sigma=1$ for unit width wave profiles,
$q_0=0$ since we do not consider
rotating black holes, and $n=2$ to allow maximum grid resolution
over the angular variations.

From the equatorially symmetric examples in
Refs. \cite{Bernstein,Brandt_ivp,Anninos_ah},
it is known that increasing the amplitude parameter $Q_0$
increases the strong field coupling of the Brill wave
and black hole, substantially distorts the spacetime
from spherical symmetry, and emits a greater fraction
of the ADM mass in the form of gravitational radiation.
To demonstrate the degree by which a black hole
is distorted from sphericity, we first look at the
geometric characteristics of the spacetime, namely the apparent
horizon since it can easily be found in the spacelike slices.
The horizon shape parameters and flat space embeddings are evaluated
for the case $\eta_0=0$ in which the Brill wave
is placed directly on the black hole throat for maximum
horizon distortion.
We use a Newton--Raphson \cite{Cook_phd,Bernstein_phd}
procedure to solve the nonlinear
equation defining the trapped surface conditions
(zero expansion of outgoing null normals to the 2--surface).
The geometric properties are extracted from the two--dimensional
sub--metric induced on the horizon surface
\begin{equation}
dl^2 = \Psi^4 \left\{\left[B+A\left(\frac{d h}{d\theta}\right)^2
       \right] d\theta^2 + D d\phi^2 + 2F d\theta d\phi \right\} ,
\label{eqn:hor}
\end{equation}
where $h(\theta)$ is the radial coordinate defining the horizon.
Visual representations of the horizon are achieved by
embedding the 2--surface given by (\ref{eqn:hor}) in a higher
three--dimensional flat space. Introducing a new coordinate $z$
on a flat 3--metric, the 2--metric of the horizon surface
is identified as 
\begin{equation}
dz^2 + d\rho^2 +\rho^2 d\phi^2 = B'(\theta) d\theta^2 + D'(\theta) d\phi^2 ,
\end{equation}
where $B'=\Psi^4(B+A(d h/d\theta)^2+F^2/D)$
and $D'=\Psi^4 D$ are the metric components of the horizon
surface transformed to a diagonal form.
Solving for the coordinate $z$ gives
\begin{equation}
z = \int d\theta \sqrt{B' - (\partial_\theta \sqrt{D'})^2} ,
\label{eqn:embed}
\end{equation}
which is integrated numerically
to obtain the embedding functions $z(\theta)$ and 
$\rho(\theta)=\sqrt{D'}$,
although an embedding is not in general guaranteed to exist.

\begin{figure}
\epsfysize=3.2in \epsfbox{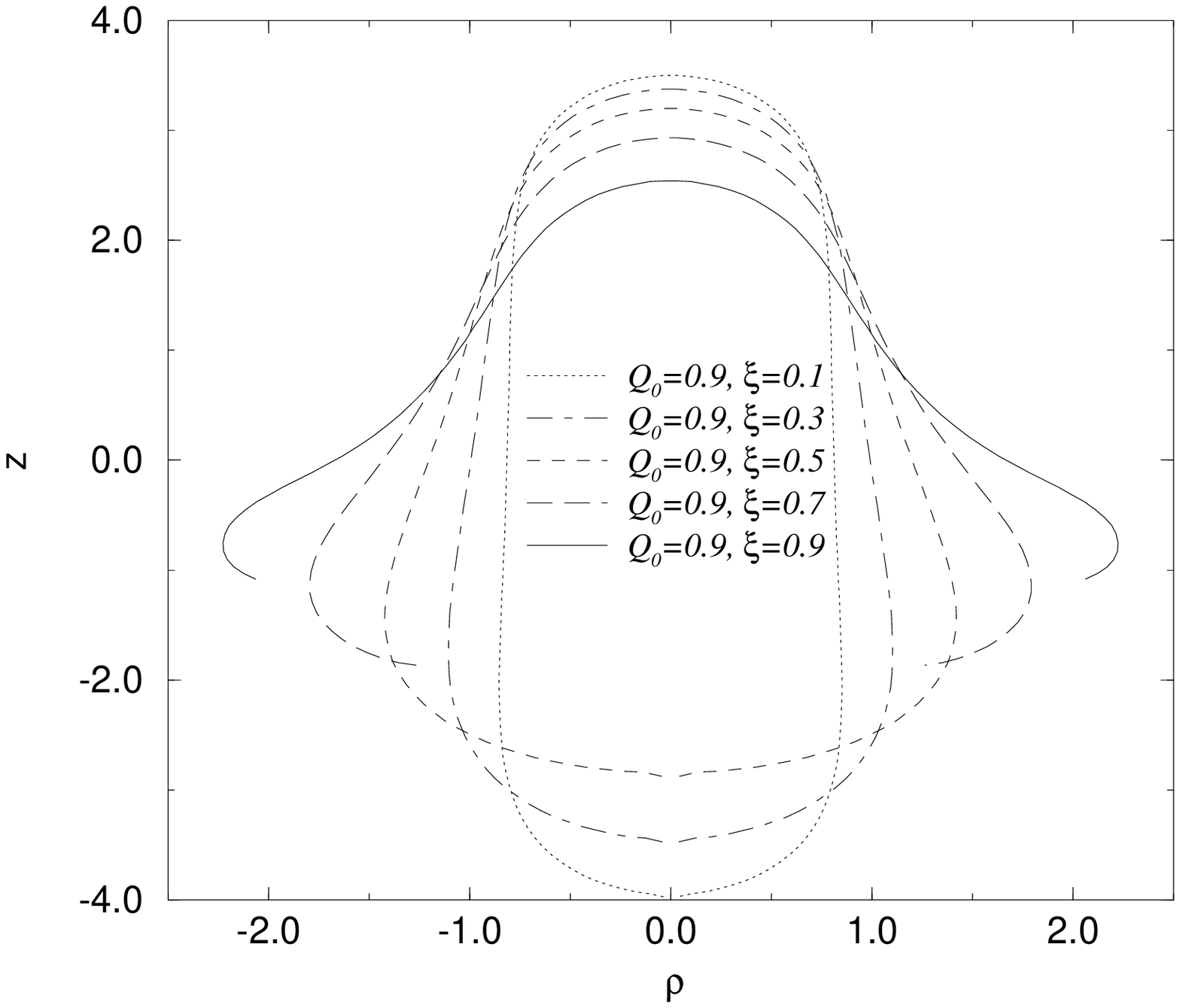}
\caption{
Embeddings in 3--dimensional flat space
of the apparent horizon 2--surfaces at the initial time
for even parity perturbations.
Each curve represents the 
$\rho$--$z$ embedding for the same strongly
distorted case with Brill wave
amplitude $Q_0=0.9$ and wave location $\eta_0=0$, but
with different mode parameters $\xi$ that regulate the
relative mixture of even and odd mode spectral distributions.
Distortions range from $R_r = 1.9$ for the 
$\xi=0.9$ case to $R_r = 4.9$ for $\xi=0.1$,
where $R_r=\mbox{max}(\sqrt{\rho^2+z^2})/\mbox{min}(\sqrt{\rho^2+z^2})$ 
is the ratio of the maximum to minimum
distances of the embedding curves to the origin
at $\rho=z=0$.
}
\label{fig:embed_q0.9}
\end{figure}
\begin{figure}
\epsfysize=3.2in \epsfbox{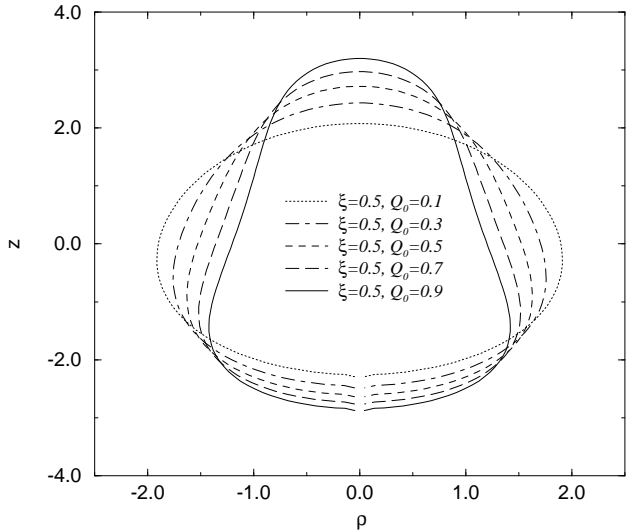}
\caption{
As figure \protect{\ref{fig:embed_q0.9}}, except here each
curve represents the embedding for the same mode parameter
$\xi=0.5$ which generates the greatest recoil effect for a
specified amplitude $Q_0$.
Distortions range from $R_r=1.2$ for
$Q_0=0.1$ to $R_r=2.7$ for $Q_0=0.9$.
}
\label{fig:embed_xi0.5}
\end{figure}

In figure \ref{fig:embed_q0.9} we show embeddings of the 
horizon 2--surface in the initial data 
(where distortions are greatest with $\eta_0=0$) 
for a highly perturbed
$Q_0=0.9$ case with different mode parameters $\xi$.
Embeddings of the more prolate odd mode distortions,
i.e. $\xi \ge 0.7$, are undefined (and therefore not displayed)
for the more negative values of $z$ due to the radical in
equation (\ref{eqn:embed}) which becomes negative.
Figure \ref{fig:embed_xi0.5} shows the equivalent embeddings for
$\xi=0.5$ as a function of wave 
amplitude $Q_0$. These embedded distortions
eventually damp out in an oscillatory fashion over time
as the horizon evolves towards sphericity after the dynamic
component has either been absorbed by the black hole, or has
propagated to asymptotic infinity in the form of gravitational waves.
Together, the embedding diagrams
indicate that horizon distortions are roughly spherical
for small amplitude perturbations and 
become generally more prolate as
$Q_0$ is increased. The shape of distortions also varies with
the mode parameter $\xi$, which regulates 
changes from equatorially symmetric
even mode behavior, to predominantly asymmetric odd mode configurations
for the larger values of $\xi$. The ratio of polar
to equatorial circumferences of the horizon surfaces, used 
in our previous work as a measure of distortion,
is not especially informative regarding the magnitude of radiation recoil.
Indeed, the purely even symmetric cases generally give rise
to greater distortions, but no recoil which is 
a function of the relative mixture of even and odd modes,
as well as the perturbation amplitude.
Assuming a simple definition of radial
distortion in the embeddings as
$R_r = \mbox{max}(\sqrt{\rho^2+z^2}) / \mbox{min}(\sqrt{\rho^2+z^2})$,
the displayed distortions range from 
$R_r = 1.2$ for ($Q_0=0.1$, $\xi=0.9$),
to 7.5 for ($Q_0=1.2$, $\xi=0.1$).

\begin{figure}
\epsfysize=3.2in \epsfbox{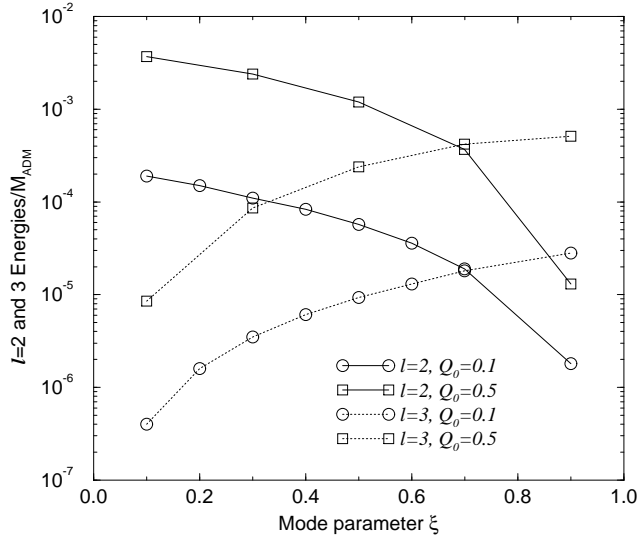}
\caption{
Radiated energies (normalized to the ADM mass, $M_{ADM}$) 
in the most dominant $\ell=2$ and 3 even parity modes as a function
of the Brill wave mode parameter $\xi$ for $\eta_0=0$ and
select values of $Q_0$.
The wave extractions are performed
on spherical shells of radius 15$M_{ADM}$ and centered
on the black hole throat.
}
\label{fig:energy}
\end{figure}
\begin{figure}
\epsfysize=3.2in \epsfbox{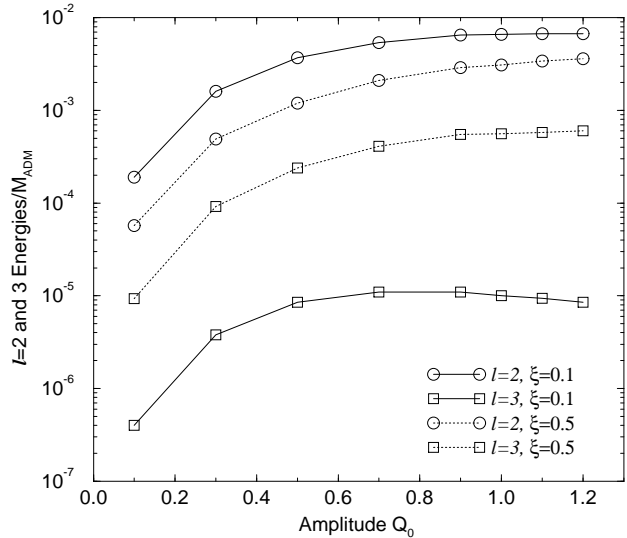}
\caption{
As figure \protect{\ref{fig:energy}}, 
except the energies are plotted
as a function of Brill wave amplitude for select values of $\xi$.
The upper bound on the wave amplitudes considered here
is dictated by the accuracy with which
the weaker of the $\ell=2$ or 3 modes can be resolved.
}
\label{fig:energy_xi0.5}
\end{figure}

The transition from even to odd mode behavior observed in the near--field
horizon embedding diagrams of figure \ref{fig:embed_q0.9},
is also mirrored by the mode distribution in the far--field radiation zone.
Figures \ref{fig:energy} and \ref{fig:energy_xi0.5} show the energy 
(normalized to the ADM mass of the spacetime) radiated in the
most dominant $\ell=2$ and 3 modes 
for $\eta_0=0$ as a function of $\xi$
and $Q_0$. As the mode parameter is
increased in figure \ref{fig:energy}, the energy distribution dominance
changes from even to odd, consistent with the horizon embeddings 
in figure \ref{fig:embed_q0.9}. Figure
\ref{fig:energy_xi0.5}
indicates that the total radiated energies asymptotically
approach constant maximal values for each of the mode parameters,
and that the range of parameters we have investigated are reasonably
representative of the most efficient radiators of gravitational energy.
We have restricted current studies to wave amplitudes
$Q_0 \le 1.2$, since the numerical results
are less reliable for larger amplitudes, 
especially at late times and in the ability
to resolve both the $\ell=2$ and 3 modes
in the extreme odd or even $\ell$-mode dominated evolutions.

The mixing of adjacent multipole modes gives
rise to a non-vanishing flux of linear momentum along the
$z$--axis which can be evaluated from products of
consecutive Zerilli wave functions \cite{Andrade}
\begin{equation}
\frac{dP^z}{dt} = \frac{1}{16\pi} \sum_{l=2}^{\infty}
                  \sqrt{\frac{(\ell-1)(\ell+3)}{(2\ell+1)(2\ell+3)}}
                  \frac{d\psi_\ell}{dt}\frac{d\psi_{\ell+1}}{dt} ,
\label{eqn:series}
\end{equation}
where the Zerilli functions $\psi_{\ell}$ are normalized
such that the total radiated energy in each mode is given by
\begin{equation}
E_{\ell}=\frac{1}{32\pi} \int dt (\dot\psi_{\ell})^2 .
\end{equation}
For numerically practical purposes, we compute only the
most significant (2,3) and (3,4) contributions. In general we find
that the higher order terms in the series (\ref{eqn:series})
play an increasingly greater role
as $\xi$ is increased and as the distortions become
dominantly odd functions. For an interesting large amplitude case 
($Q_0 = 0.9$) the momentum ratio
$P_{(2,3)} / P_{(3,4)}$
varies from roughly 50 to 0.2 for
$\xi=0.1$ and 0.9 respectively, with even greater
ratios for the smaller amplitude cases.
However, as we show below, the
greatest recoil velocities arise for roughly equal mixtures
of even and odd perturbations (i.e., $\xi=0.5$), and for these
cases the (2,3) contribution exceeds the (3,4) by 
at least a factor of 50 in all cases we have studied.
The results presented for the radiated momentum are derived by summing
both the (2,3) and (3,4) contributions. To confirm the degree to which
these two dominant pairs are complete, and to independently check
our calculations, we have also evaluated
the momentum flux from the Landau--Lifshitz pseudotensor
\cite{Landau,Cunningham_odd,Cunningham_even}
\begin{equation}
\frac{d P^z}{dt} = \frac{r^2}{16\pi} \int \left[ (\dot F)^2
                 +\frac{1}{4} (\dot B - \dot D)^2\right]
                 \cos\theta~d\Omega ,
\label{eqn:landau}
\end{equation}
which is valid for both even ($F=0$) and odd ($B=D=0$)
parity perturbations. We find excellent
agreement, typically better than 10\%, between the two calculations.

\begin{figure}
\epsfysize=3.2in \epsfbox{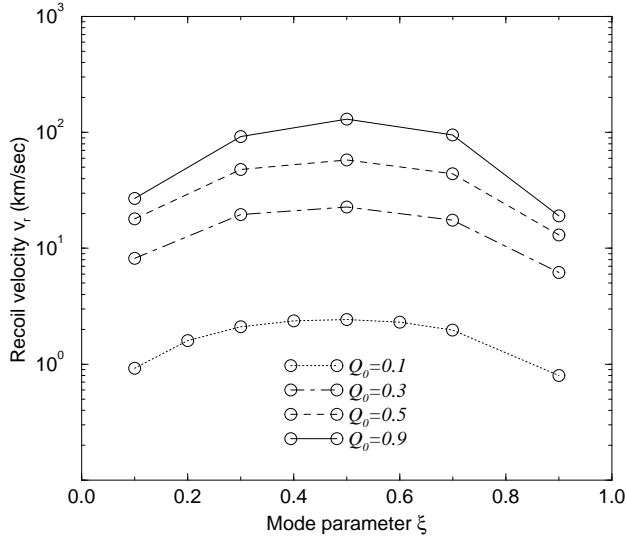}
\caption{
Recoil velocities for even parity distortions
as a function of mode parameter $\xi$ for $\eta_0=0$ and
different wave amplitudes $Q_0$.
The velocities are computed from the radiative momentum flux
carried by gravitational waves crossing spherical shells of
radius 15$M_{ADM}$ from the black hole throat.
}
\label{fig:recoil_xi}
\end{figure}
\begin{figure}
\epsfysize=3.2in \epsfbox{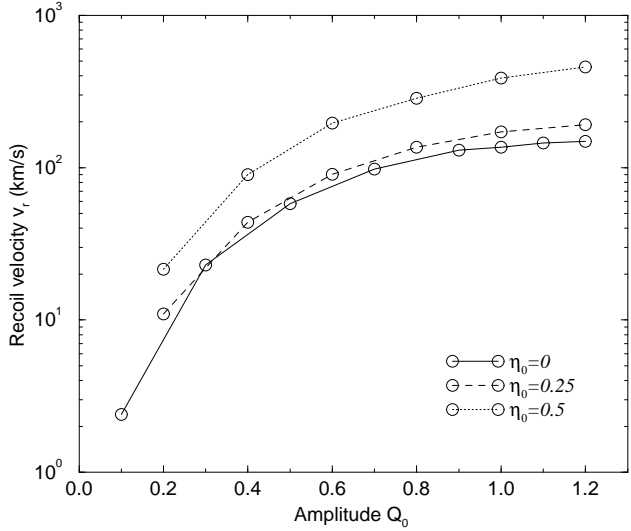}
\caption{
Recoil velocities for even parity perturbations as a function of 
Brill wave amplitude $Q_0$ for $\xi=0.5$ and
different initial wave locations $\eta_0$.
The trend toward asymptotic flatness in the displayed curves suggests
maximum recoil velocities of about 150, 200 and 500 km/sec,
for $\eta_0$ = 0, 0.25 and 0.5 respectively.
}
\label{fig:recoil_q0}
\end{figure}

As a result of the momentum emission from the interplay
between even and odd $\ell$--modes, the final black hole
will acquire a recoil velocity
\begin{equation}
v^z_r = - \frac{1}{M_{ADM}} \int \left(\frac{dP^z}{dt}\right) dt ,
\end{equation}
that is opposite in direction to the momentum flux of the waves.
We show these velocities in figures \ref{fig:recoil_xi}
and \ref{fig:recoil_q0} in physical units of kilometers per second,
and as a function of the parameters $\xi$, $\eta_0$ and $Q_0$.
Figure \ref{fig:recoil_xi} confirms that maximum recoil
occurs at $\xi=0.5$. The asymptotic flatness of the curves
at large values of $Q_0$ in figure \ref{fig:recoil_q0}
suggests maximum velocities of about 150, 200, and 500 km/sec
for the most strongly distorted cases that we are able to compute
numerically for $\eta_0$ = 0, 0.25 and 0.5 respectively. 
However, we are unable to reliably evolve greater amplitudes for 
$\eta_0$ = 0.25 or 0.5
and establish a precise turnover velocity since
our code breaks at these large amplitudes, though the curves already
show signs of flattening by $Q_0=1.2$.
Hence the quoted values in these two cases are approximate extrapolations.
Also notice that greater recoil velocities result when the Brill waves
are placed at larger radii, beyond the 
perturbation potential barrier (at $r>3M$, or equivalently $\eta>1.8$). 
In these cases, the 
ingoing waves excite the ringing modes more
strongly as they cross the barrier and emit a greater flux of
energy--momentum. On the other hand, it is also likely that
a large fraction of the radiated flux 
in the large $\eta_0$ cases can be attributed
to the outgoing wave component and general distortions
of the global spacetime 
\cite{Bernstein_phd}, and not to pure black hole ringing from
localized collapse or impact scenarios.

\subsection{Odd Parity}
\label{subsec:odd_schw}

Following the general presentation of \S\ref{subsec:even_schw},
we present in this section recoil velocities from odd
parity distortions of black holes as a function of mode parameter ($\xi$),
initial peak location ($\eta_0$),
and wave amplitude ($Q_0$) in the initial data of \S\ref{sec:initial}.

The momentum arising from consecutive $\ell$--mode interactions
of odd parity waves takes a form analogous to (\ref{eqn:series}),
except the wave functions $\psi_{\ell}^{odd}$ 
(replacing the even parity $\dot\psi_{\ell}$) are extracted from 
the $\gamma_{\theta\phi}$ metric component by
\begin{equation}
\psi_{\ell}^{{odd}} = \sqrt{\frac{2(\ell - 2)!}{(\ell+2)!}}
                    \int \partial_\eta F
                    (\partial_\theta^2 - \cot\theta \partial_\theta)
                    Y_{\ell 0} d\Omega .
\end{equation}
In this form, the wave functions
are related to the Regge-Wheeler \cite{RW57} perturbation variable $h_2$
\begin{equation}
\psi_{\ell}^{{odd}} 
                  = \sqrt{\frac{(\ell+2)!}{2(\ell-2)!}}
                    ~r\partial_r \left(\frac{h_2}{r^2}\right) ,
\end{equation}
with normalization
\begin{equation}
E_\ell = \frac{1}{32\pi} \int dt (\psi_{\ell}^{odd})^2 .
\end{equation}

\begin{figure}
\epsfysize=3.2in \epsfbox{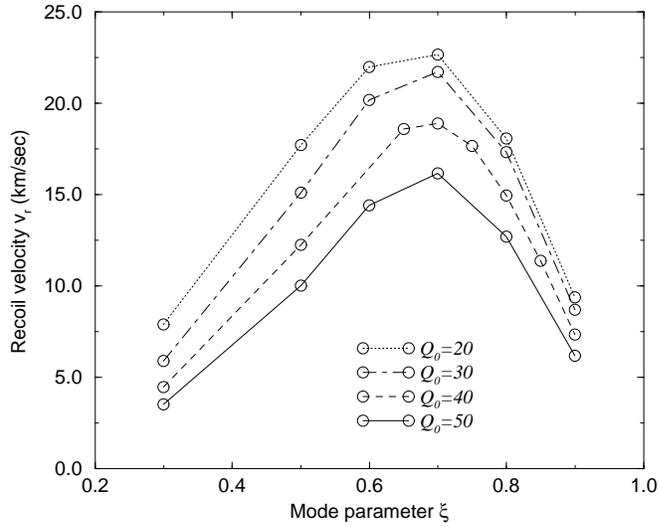}
\caption{
Recoil velocities for odd parity distortions
as a function of mode parameter $\xi$ for $\eta_0=0$ and
different wave amplitudes $Q_0$.
}
\label{fig:odd_xi}
\end{figure}

\begin{figure}
\epsfysize=3.2in \epsfbox{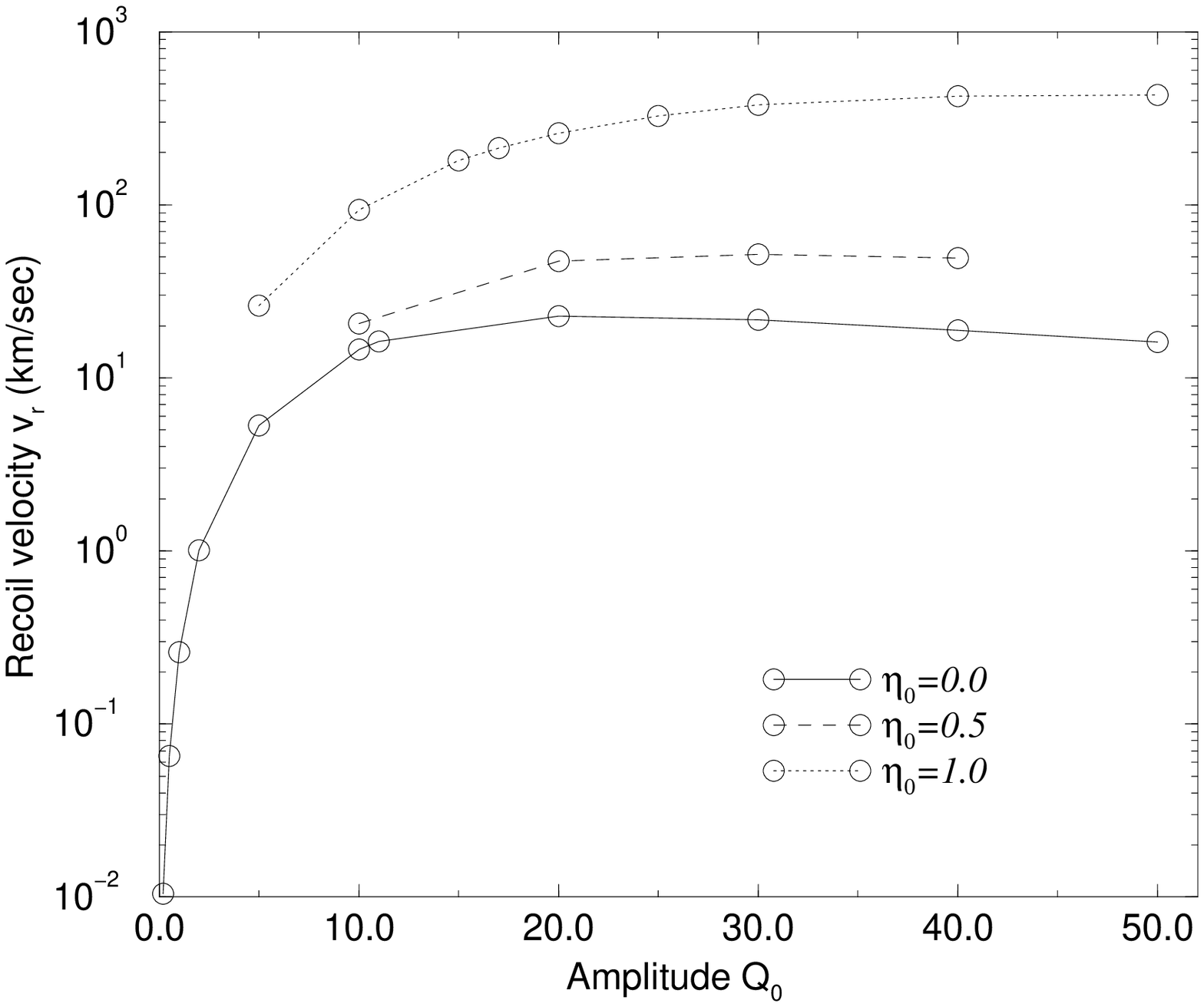}
\caption{
Recoil velocities for odd parity perturbations as a function of
Brill wave amplitude $Q_0$ for $\xi=0.7$ and
different initial wave locations $\eta_0$.
Maximum recoil velocities of about 23, 52 and 430 km/sec are
found for $\eta_0$ = 0, 0.5 and 1 respectively.
}
\label{fig:odd_q0}
\end{figure}

An important difference in the evolutions of odd (versus even) parity
data is that it is necessary to keep track of the momentum
contributions from a greater number of
$\ell$--mode pairs since there
is no clear dominance by the lowest order terms. The individual
contributions can jump from positive to negative values of momentum
with roughly the same amplitude
over consecutive pairs. For example, the better
behaved $\eta_0=0$ small amplitude cases require at least three
pairs in the series to converge at the 10\% level when
compared with subsequent additions and to the 
Landau--Lifshitz formula (\ref{eqn:landau}). 
Also, the odd parity initial value problem and the
dynamical evolutions over time can both generate
significant even parity signals, contributing up to 
a few percent of the net recoil velocity in the cases
we have investigated. Hence all results in this section
are derived from the Landau--Lifshitz pseudotensor, thus accounting
for all the modes in
evaluating the velocities (though we have compared and confirmed the 
consistency of both methods for small and large amplitude runs).
We plot in Figures \ref{fig:odd_xi}
and \ref{fig:odd_q0} the velocities as a function of
$\xi$ and $Q_0$ respectively. Figure \ref{fig:odd_xi}
indicates that maximum recoil is achieved for $\xi\approx 0.7$.
Figure \ref{fig:odd_q0} shows the
maximum velocity for $\xi=0.7$ as $Q_0$ is varied over
a numerically robust range of amplitudes for three
different initial wave positions
$\eta_0$ = 0, 0.5 and 1. The evolutions generate maximum velocities
of 23, 52 and 430 km/sec, with increasingly greater
velocities for Brill waves initially concentrated at greater 
distances from the black holes.

In comparing these results with the even parity cases
(say $\eta_0$ = 0),
one should not necessarily conclude that odd parity radiation
is less effective in producing radiation recoil for any intrinsic
reason.  Much has to do with the manner in which the data was constructed.
For example,
the even parity data distorts the spatial metric, while the odd parity
data uses a conformally flat metric.  One could, in principle, produce
even parity distortions of the spacetime through the extrinsic curvature
while maintaining a flat metric.  Likewise, one could add odd parity
data to the metric itself rather than the extrinsic curvature.  This
would make the procedures more similar (and possibly the radiation
energies as well).  Furthermore, there are many
ways to construct initial data for both types of radiation and it
is not feasible to study them all.  Rather, our results represent the
maximal effects of a certain class of black hole distortions.

\section{Conclusion}
\label{sec:summary}

We have carried out a systematic study of single black holes
distorted by strong--field axisymmetric Brill waves in an effort to
quantify the astrophysical significance of the ``rocket'' effect imparted
to the final black hole from the momentum carried by gravitational
radiation in the system. This work compliments our previous studies
of the head--on collision of two unequal mass black holes
\cite{Anninos1}, where we found recoil velocities up to
10--20 km/sec. However, it is likely that coalescing binary black
holes with arbitrary physical parameters (i.e., impact parameters,
masses, and spins) may generate greater recoil velocities, so we have
focused these current studies to deduce the maximum recoil expected from
highly asymmetrical configurations. The Brill wave + black hole systems
we have studied allow a parameterization of the wave strengths,
widths, locations, and shapes of the perturbing sources
such that we are able to systematically explore the role of
various parameters in fully nonlinear numerical calculations of
strongly distorted black hole spacetimes. With this approach, we are able 
to generate greater distortions and wider spectral
energy distributions of black holes than
observed in our simulations of colliding binary systems.
We thus also consider our current results as reasonable
maximum estimates of radiation recoil in
single or late--stage binary black hole systems (although a more
precise comparison between single and binary evolutions must also
account for any residual radiation content in the initial data
of the respective systems).

For the most highly distorted spacetimes, we find
maximum recoil velocities in excess of
400 km/sec for both even and odd parity data with Brill waves
initially centered at large distances from the black hole throat,
e.g., $\eta_0=0.5$ (1.0) for even (odd) parity perturbations.
Our results exhibit a strong dependence on the initial
placement of the Brill waves, as well as their amplitude
and spectral composition. Of all these effects, we are
less certain of the role which the initial wave placement $\eta_0$
plays in generating a true maximum value, since for the
numerically difficult combination of large
separations and amplitudes, our code eventually breaks down.
However, we expect for radiation clumps located further
from the black hole, that a substantial fraction of the 
(outgoing component of the) Brill waves escapes
to infinity since the perturbations are applied essentially to the spacetime
surrounding the black hole, and not directly on the throat. Hence we
expect that the bulk of emitted energy--momentum flux can be attributed to
the initial wave configuration for large $\eta_0$, as opposed to any intrinsic
ringing of the black hole associated with localized source dynamics
such as ingoing wave collisions, 
collapsing stellar cores, or coalescing binaries.
On the other hand, an ingoing wave located outside
the potential barrier can scatter off and impart a much
greater momentum to the hole.  We were, however, unable to
distinguish secondary wave pulses in our numerical data corresponding
to reflected waves.

In addition to reducing the initial outgoing Brill wave content,
it is also likely that the $\eta_0=0$
cases represent more appropriate late stage recoil models for
black hole binary systems. In these cases, our
results of 150 and 23 km/sec for even and odd parity perturbations
are in general agreement with the bound
$v_r \le 300$ km/sec derived by Bekenstein \cite{Bekenstein}
in his quasi--Newtonian considerations of 
the interaction between quadrupole and octupole terms in non-spherical
stellar core collapse to black holes. Furthermore, the odd parity
recoil in our calculations is remarkably similar to
the 25 km/sec found by Moncrief \cite{Moncrief} for
non-spherical models of black hole formation.
Our even parity results
are approximately a factor of two times
larger than the quasi--Newtonian calculations ($v_r \sim 67$ km/sec) of
binary systems in Keplerian orbits by Fitchett \cite{Fitchett1}.
However, considering the ambiguity in choosing the
final prior-to-plunging orbit and in extrapolating the 
perturbation calculations
to the equal mass limit, our results
are in fairly good agreement with the predictions
of Fitchett and Detweiler \cite{Fitchett2} who 
extended Fitchett's earlier work to perturbation theory and computed
a maximum velocity of about 120 km/sec for the merging of two
black holes from the last stable circular orbit.
We are also in agreement with 
Nakamura, Oohara and Kojima \cite{Nakamura} who estimate
a maximum velocity of about 240 km/sec
from numerical perturbative calculations of test bodies
plunging into black holes from infinity with arbitrary orbital
angular momentum.

In comparison, the escape velocity from galactic structures
can vary from about several hundred km/sec for spiral galaxies such
as the Milky Way, to about a thousand km/sec for the more massive
giant ellipticals such as M87. Our results, however, suggest
that black holes which may be located in
the centers of galaxies and which undergo highly asymmetric
evolutions (including strong field distortions and binary mergers)
are relatively stable entities 
and will not likely escape from the host galaxy, assuming that
the ``on the throat'' numerical calculations
are reasonably representative models.
Although we have established that the recoil effect is not
generally large enough to be considered astrophysically significant,
this does not, however, rule out the possibility of black hole
ejections from galactic disks far from the core and in
the direction of galactic rotation, nor the
possibility of black hole ejections from globular cluster
systems in galactic halos. 
Black holes can more readily escape from these systems
to wander through the galaxy or even intergalactic space.

\acknowledgements
The numerical simulations were performed on the Origin2000
machines at the NCSA and the Albert Einstein Institute.
This work was supported by NSF grant PHY 98-00973, and
performed under the auspices of the U.S. Department
of Energy by Lawrence Livermore National Laboratory under Contract
W-7405-Eng-48.


\begin{thebibliography}{0}
\bibitem{Bonnor}
Bonnor and Rotenberg, Proc. Roy. Soc. {\bf A265}, 109 (1961)

\bibitem{Perez}
A. Perez, Phys. Rev. {\bf 128}, 2471 (1962)

\bibitem{Bekenstein}
J.D. Bekenstein, Astrophys. J. {\bf 183}, 657 (1973)

\bibitem{Moncrief}
V. Moncrief, Astrophys. J. {\bf 238}, 333 (1980)

\bibitem{Fitchett1}
M.J. Fitchett, MNRAS {\bf 203}, 1049 (1983)

\bibitem{Fitchett2}
M.J. Fitchett and S. Detweiler, MNRAS {\bf 211}, 933 (1984)

\bibitem{Nakamura}
Nakamura, Oohara and Kojima, Prog. Theor. Phys. Suppl. {\bf 90}, 135 (1987)

\bibitem{Andrade}
Z. Andrade and R.H. Price, Phys. Rev. D, {\bf 56}, 6336 (1997)

\bibitem{Anninos1}
P. Anninos and S. Brandt,
Phys. Rev. Lett., {\bf 81}, 508 (1998)

\bibitem{Brill}
D.S. Brill, Ann. Phys. (N.Y.) {\bf 7}, 466 (1959)

\bibitem{Bernstein}
D. Bernstein, D. Hobill, E. Seidel and L. Smarr,
Phys. Rev. D, {\bf 50}, 3760 (1994)

\bibitem{Brandt_ivp}
S. Brandt and E. Seidel,
Phys. Rev. D, {\bf 54}, 1403 (1996)

\bibitem{Anninos_ah}
P. Anninos, D. Bernstein, S. Brandt, D. Hobill, 
E. Seidel and L. Smarr,
Phys. Rev. D, {\bf 50}, 3801 (1994)

\bibitem{Anninos_2bh1}
P. Anninos, D. Hobill, E. Seidel, L. Smarr and W.M. Suen,
Phys. Rev. Lett., {\bf 71}, 2851 (1993)

\bibitem{Anninos_2bh2}
P. Anninos, D. Hobill, E. Seidel, L. Smarr and W.M. Suen,
Phys. Rev. D, {\bf 52}, 2044 (1995)

\bibitem{Bowen}
J. Bowen and J.W. York,
Phys. Rev. D, {\bf 21}, 2047 (1980)

\bibitem{Cook_phd}
G. Cook, Ph.D. thesis, University of North Carolina at Chapel Hill,
Chapel Hill, North Carolina (1990)

\bibitem{Bernstein_phd}
D. Bernstein, Ph.D. thesis, University of Illinois,
Urbana-Champaign (1993)

\bibitem{Landau}
Landau, L.D. and Lifshitz, E.M.,
{\it The Classical Theory of Fields},
(Pergamon Press, 1975)

\bibitem{Cunningham_odd}
Cunningham, C.T., Price, R.H. and Moncrief, V.,
Astrophys. J. {\bf 224}, 643 (1978)

\bibitem{Cunningham_even}
Cunningham, C.T., Price, R.H. and Moncrief, V.,
Astrophys. J. {\bf 230}, 870 (1979)

\bibitem{RW57}
T. Regge and J. Wheeler, Phys. Rev. {\bf 108}, 1063 (1957)

\end{thebibliography}
\end{document}